\documentclass[prd,10pt,aps,superscriptaddress,twocolumn,amsmath,amssymb,nofootinbib,longbibliography]{revtex4-1}

\usepackage{natbib}
\usepackage[colorlinks]{hyperref}
\hypersetup{linkcolor=blue,citecolor=blue,filecolor=blue,urlcolor=blue}

\usepackage{afterpage}
\usepackage{float} 

\usepackage{lipsum}
\usepackage{graphicx}
\usepackage{subfigure}

\usepackage{tabularx}
\usepackage{multirow}
\usepackage{array}
\usepackage[dvipsnames]{xcolor}
\usepackage{orcidlink}

\usepackage{dcolumn}
\usepackage{bm}
\usepackage{amsfonts}
\usepackage{bbm}
\usepackage{bbold}
\usepackage{relsize}
\usepackage{paralist}
\usepackage{physics}

\usepackage{wasysym}

\definecolor{green}{rgb}{0.19,0.64,0.54}
\definecolor{blue}{rgb}{0,0,1}
\definecolor{reddish}{rgb}{0.65, 0.2, 0.2}
\definecolor{darkgreen}{rgb}{0.2,0.7,0.3}
\definecolor{darkblue}{rgb}{0.3,0.40,0.48}
\definecolor{gray}{rgb}{.8,.8,.8}

\newcommand{\Icut}{%
  \mathrel{\vcenter{\offinterlineskip
  \hbox{$I$}\vskip-0.9ex\hbox{\scalebox{0.9}{--}}}}}

\begin{document}

\title{Gravitational wave signals from primordial black holes orbiting
  solar-type stars}

\author{Vitorio A.~\surname{De Lorenci} \orcidlink{0000-0001-5880-2207}}
\email{delorenci@unifei.edu.br}

\affiliation{Instituto de F\'{\i}sica e Qu\'{\i}mica, Universidade
  Federal de Itajub\'a, Itajub\'a, Minas Gerais 37500-903, Brazil}

\author{David I.~\surname{Kaiser} \orcidlink{0000-0002-5054-6744}}
\email{dikaiser@mit.edu}

\affiliation{Department of Physics, Massachusetts Institute of
  Technology, Cambridge, Massachusetts 02139, USA}

\author{Patrick \surname{Peter} \orcidlink{0000-0002-7136-8326}}
\email{peter@iap.fr}

\affiliation{{${\cal G}\mathbb{R}\varepsilon\mathbb{C}{\cal
      O}$}---Institut d'Astrophysique de Paris, CNRS and Sorbonne
  Universit\'e, UMR 7095 98 bis Boulevard Arago, 75014 Paris, France}

\author{Lucas S.~\surname{Ruiz} \orcidlink{0000-0002-5705-5278}}
\email{lucasruiz@unifei.edu.br}

\affiliation{Instituto de Matem\'atica e Computa\c{c}\~ao,
  Universidade Federal de Itajub\'a, \\ Itajub\'a, Minas Gerais
  37500-903, Brazil}

\author{Noah E.~\surname{Wolfe} \orcidlink{0000-0003-2540-3845}}
\email{newolfe@mit.edu}

\affiliation{Department of Physics, Massachusetts Institute of
  Technology, Cambridge, Massachusetts 02139, USA}

\affiliation{LIGO Laboratory, Massachusetts Institute of Technology,
  185 Albany St, Cambridge, MA 02139, USA}


\begin{abstract}
Primordial black holes (PBHs) with masses between $10^{14} $ and
$10^{20}$ kg are candidates to contribute a substantial fraction of
the total dark matter abundance. When in orbit around the center of a
star, which can possibly be a completely interior orbit, such objects
would emit gravitational waves, as predicted by general relativity.
In this work, we examine the gravitational wave signals emitted by
such objects when they orbit typical stars, such as the Sun. We show
that the magnitude of the waves that could eventually be detected on
Earth from a possible PBH orbiting the Sun or a neighboring Sun-like
star within our galaxy can be significantly stronger than those
originating from a PBH orbiting a denser but more distant neutron star (NS). Such signals may
be detectable by the \textsmaller{LISA} gravitational-wave detector. In addition, we
estimate the contribution that a large collection of such PBH-star
systems would make to the stochastic gravitational-wave background
(SGWB) within a range of frequencies to which pulsar timing arrays are
sensitive.
\end{abstract}
		
\maketitle

\section{Introduction}
\label{introduction}

Primordial black holes (PBHs) are hypothetical black holes that could
have formed in the very early universe, for example from the
gravitational collapse of primordial perturbations that were amplified
during a phase of inflation.  Although no PBHs have been
conclusively detected, they can act as a component of dark matter. The relevant constraints leave open a window
of masses $10^{14} \, \text{kg} \leq m \leq 10^{20} \,
\text{kg}$, often dubbed the ``asteroid-mass range,'' within which PBHs could constitute the entire dark matter
abundance~\cite{Carr:2020xqk,Green:2020jor,Carr:2023tpt,Escriva:2022duf,
  Gorton:2024cdm,Khlopov:2024nqp}.

The discovery of gravitational waves (GWs) has opened up an entirely
new branch of astronomy. Since the first detection, there have been
numerous other detections of GWs from black hole mergers and neutron
star collisions, significantly enhancing our understanding of the
universe \cite{Bailes:2021tot}. In this paper we study several scenarios in which PBHs within the asteroid-mass range could yield observable GW signals.

Typically when possible GW signals from asteroid-mass PBHs have been considered, the focus has been on primordial tensor perturbations induced at second order in perturbation theory from the large-amplitude scalar curvature perturbations that would have undergone gravitational collapse at very early times to yield a population of PBHs. The peak frequency of such primordial GWs depends on the typical mass $m$ with which the PBHs form; for PBHs in the asteroid-mass range, such induced GW signals today would peak in the range $f \sim {\cal O} (10^0 - 10^3) \, \text{Hz}$, with predicted amplitudes to which upcoming GW detectors, such as the Einstein Telescope \cite{Abac:2025saz} and Cosmic Explorer \cite{Reitze:2019iox}, should be sensitive. (See, e.g., Refs.~\cite{Domenech:2021ztg,Qin:2023lgo}.)

In this paper we consider GWs arising from very different processes involving asteroid-mass PBHs, with correspondingly different frequencies and hence distinct opportunities for detection in upcoming detectors. In particular, we build upon previous work in which trajectories of small primordial
black holes bound to stellar objects were studied in detail
\cite{2024arXiv240508113D}, and calculate the
characteristic features of the GWs that such systems should emit. For
a PBH of mass $m \simeq 10^{20} \, \text{kg}$ trapped in
the Sun, we show that the resulting GWs from the PBH's orbital motion should be detectable in future GW
experiments such as \textsmaller{LISA} \cite{LISA:2022yao}, since the very faint amplitude at emission would be compensated by the very
short distance between our star and the detectors. (Compare with Refs.~\cite{2024arXiv240201838B,2024arXiv240408735B,2024PhRvD.109f3004B}.)

In addition to considering the GW signals from single PBH-star
systems, we also estimate the contribution that a large collection of
such systems would make, integrated over cosmic history, to the
stochastic gravitational-wave background (SGWB). Recent measurements
of the SGWB using pulsar timing arrays are in tension with predictions
of the signal that would arise from the presumed dominant
contribution, namely, the binary inspiral of supermassive black holes
\cite{NANOGrav:2023hvm}. We find that a large collection of small-mass
PBHs orbiting ordinary Sun-like stars would contribute to the SGWB and
could help ease the present tension with observations.

In Section \ref{model} we introduce our parameterization for the
PBH-star systems, and in Section \ref{sec:GWtheory} we identify the
dominant (quadrupole) contribution to the GW emission from such
systems. We also identify appropriate time-scales within which our
estimates remain self-consistent. Section \ref{simulations} presents
results for individual GW waveforms resulting from a variety of
PBH-star orbits, including those in which the PBH remains entirely
bound within its host star. We then turn in Section
\ref{sec:GWlocalsource} to study whether such individual-system GW
signals might be detectable with upcoming GW experiments, such as
\textsmaller{LISA}. In Section \ref{sec:SGWB} we estimate the expected contribution
to the SGWB, including the likely spectral index from a large
collection of such PBH-star systems over cosmic history. We present
concluding remarks in Section \ref{sec:Final}.

\section{Dynamics of a PBH in a stellar orbit}
\label{model}

Following the methodology in Ref.~\cite{2024arXiv240508113D}, we
consider the dynamics of a PBH of mass $m$ in orbit around the
center of a star of mass $M_{\star}$ and radius $ R_\star$. (See
Fig.~\ref{Config} for the geometric configuration discussed below.)
The PBHs of interest have very small cross-sections, due to their
asteroid-sized masses, so that their orbits can occur partially or
even totally inside the star interior ($r< R_\star$). In this case, the
gravitational potential energy $U(r)$ of the PBH will depend on its
radial position in such a way that
\begin{equation}
U(r) = \int_{+\infty}^{r} \frac{G m M(u) }{u^2} \dd u,
\end{equation}
where
\begin{equation}
M(r) = 
\begin{cases} 
\displaystyle 4\pi \int_{0}^r \rho(v) v^2 \dd v & \text{for } r<R_\star, \\
& \\
M_\star, & \text{for }  r\geq R_\star,
\end{cases}
\end{equation}
represents the enclosed star mass that effectively interacts with the
PBH when it is at a distance $r$ from the origin.

The motion of a particle under such a potential preserves its angular
momentum $L= m r^2 \dot{\varphi} \doteq m \ell$, and also its total
energy $E_\textsc{t} = \frac12m \dot{r}^2 + \frac12m r^2 \dot{\varphi}^2 +
U(r)$, and $U(r)$ coincides with the Keplerian potential energy $U(r)
=-G m M_\star /r$ when $r\geq  R_\star$, as expected.

The differential equation governing the orbital motion of a PBH can be
conveniently written in terms of the dimensionless radial distance $s
= r/ R_\star$ and time $\tau = \sqrt{G M_{\star}/R_\star^3} \, t$ variables as
\begin{equation}
s''  =   \frac{\bar{\ell}^2}{s^3}  -  \frac{\bar{M}(s)}{s^2},
\label{oderadius3}
\end{equation}
where a prime denotes a derivative with respect to $\tau$. The
dimensionless mass function is defined through $\bar{M}(s) =
M(s)/M_\star$, while the angular-momentum density is $\bar{\ell} =
\ell/\sqrt{G M_\star  R_\star}$.

The angular distance between the apocenter $r_\text{max}$ and
pericenter $r_\text{min}$ of the orbit is given by (see the discussion
in the appendix of Ref.~\cite{2024arXiv240508113D})
\begin{equation}
\delta \varphi = \int_{s_\text{min}}^{s_\text{max}}
\frac{\bar{\ell}}{\sqrt{2\left[ \bar{E}_\textsc{t} - \bar V(s)
      \right]}} \frac{\dd s}{s^2},
\label{deltaphi2}
\end{equation}
where $s_\text{max} = r_\text{max}/ R_\star$ and $s_\text{min} =
r_\text{max}/ R_\star$ give the maximum and minimum distances from the
particle's orbit to the origin, $\bar V(s)$ is the dimensionless
effective potential of the particle, defined through
\begin{equation}
\bar V(s)= \frac{\bar{\ell}^2}{2 s^2} + \int_{+\infty}^{s}
\frac{\bar{M}(u) }{u^2} \dd u ,
\label{vbar}
\end{equation}
and 
\begin{equation}
E_\textsc{t} = \frac{G m M_{\star}}{ R_\star}\bar{E}_\textsc{t} = \frac{G m
  M_{\star}}{ R_\star}\left [ \frac{1}{2} s'^2 + \bar V(s) \right],
\end{equation}
is the dimensionless total energy of the particle.

As discussed in Ref.~\cite{2024arXiv240508113D}, the leading
relativistic correction to the equation of motion for the PBH is given
by $3 G M (r) \ell^2 / (r^4 c^2)$. The ratio of this term to the
Newtonian $G M(r) / r^2$ term is $3 (\ell / cr)^2$, which is found to
remain below $\mathcal{O} (10^{-5})$ for the orbital motions under
study here; hence we can safely neglect post-Newtonian corrections
over the time-scales of interest. Likewise, dynamical friction for
such systems typically scales as $t_\text{dyn} \simeq 10^{-1} ( M_\star / m)
\,t_\text{orbit}$, where $t_\text{orbit} \sim R_\star / v_0$
\cite{Ostriker_1999,2024arXiv240408735B,2024arXiv240408057C}. For $M_\star
\sim M_\odot$, $R \sim R_\odot$, and $v_0 \sim {\cal O} (10^2) \,
\text{km} \, \text{s}^{-1}$, this suggests that dynamical friction
should affect the PBH's orbit on a time-scale $10^6 \, \text{yr} \leq
t_\text{dyn} \leq 10^{12} \, \text{yr}$ for PBH masses within the
range $10^{14} \,\text{kg} \leq m \leq 10^{20} \, \text{kg}$, many
order of magnitude longer than the typical orbital time
$t_\text{orbit} \sim 10^4 \, \text{s}$. We therefore also neglect
dynamical friction over the relevant time-scales for our calculations.

\section{Gravitational waves}
\label{sec:GWtheory}

\begin{figure}[t]
\includegraphics[width=\linewidth]{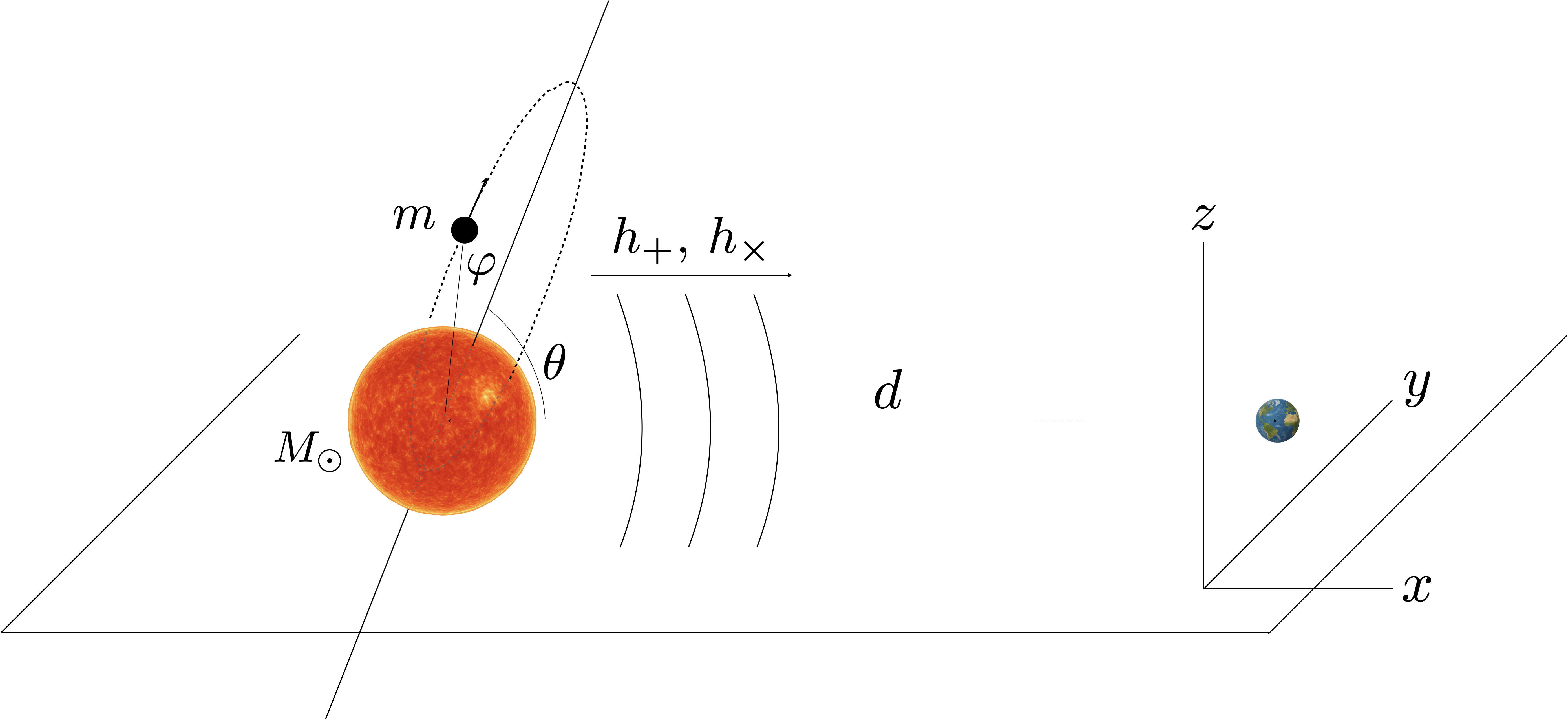}

\caption{Illustration for the Sun-Earth example of the geometric
  configuration for the emission of gravitational waves with
  polarizations $h_+$ and $h_\times$.  The trajectory of the mass $m$
  orbiting the mass $M_\odot\gg m$ lies entirely in a plane making an
  angle $\theta$ with the ecliptic plane. In the best-case scenario,
  one has $\theta=\pi/2$ so the PBH remains in the $y-z$ plane, with
  GW emission along the $x-$direction.}

\label{Config}
\end{figure}

For the parameter ranges of interest, the system we study falls within
a weak gravitational regime. In that case, the gravitational waves
will be dominated by the quadrupole moment. The next-leading
correction beyond quadrupole is the first post-Newtonian (PN)
contribution, which is expected to be of order ${\cal O}(v^2/c^2)$. In
the small-velocity regime considered here, this term introduces only a
minor correction that remains negligible over the time-scale relevant
to our analysis. In that case, the transverse-traceless (TT) component
of the gravitational wave strain tensor takes the form
\cite{1986bhwd.book.....S} \begin{equation} h_{ij}^\textsc{tt} =
  \frac{2G}{c^4 d} \perp_{ijk\ell} \ddot{\Icut}_{k\ell},
\label{shapiro1}
\end{equation}
where $d$ is the distance between the source and the observer and
$\Icut_{ij}$ denotes the traceless part of the mass quadrupole moment
tensor of the source,
\begin{equation}
\Icut_{jk} =  m \left(x_j x_k - \frac{1}{3} \bm{x}^2 \delta_{jk}  \right),
\label{shapiro2}
\end{equation}
with $x_i$ the $i$th component of the position of the PBH in its
planar orbital motion and $\bm{x}^2=\delta^{ij}x_i x_j$. In
Eq.~\eqref{shapiro1}, the projector $\perp_{ijk\ell}$ along the unit
vector $n_i$ is given by
\begin{equation}
\perp_{ijk\ell} = \perp_{(i k} \perp_{j)\ell} -\frac12 \perp_{ij}
\perp_{k\ell},
\end{equation}
with $\perp_{ij} = \delta_{ij} - n_i n_j$, and the symmetrization is
only in $i$ and $j$.

We neglect the effect of time retardation, as our interest lies only
in the magnitude of the emitted gravitational radiation. This
approximation simplifies the analysis but at the cost of losing
precise information about the timing and relative phase of the
gravitational waves, which could be significant, for instance, when
the source is rapidly changing or located far from the
observer. However, the influence of the retardation effect for
periodic or quasi-periodic orbits is generally less important, as the
regular pattern of the emission enables one to determine the frequency
and amplitude of the waves, which are the main physical
characteristics of the radiation \cite{1980RvMP...52..299T}.

With respect to the dimensionless variables, defined by $\bar x_i =
x_i/ R_\star$, the orbit of the PBH has the form, in the reference frame
defined in Fig.~\ref{Config},
\begin{equation}
(\bar{x},\bar{y},\bar{z}) = s(t) \left\{ \cos\theta, \cos \left[
\varphi(t) \right] \sin\theta, \sin \left[ \varphi(t) \right] 
    \sin\theta \right\},
\end{equation}
where $\theta$ is a constant indicating the overall inclination of the
PBH trajectory with respect to the ecliptic plane. From now on, and
for the sake of simplicity, we shall assume $\theta\to\pi/2$, thereby
maximizing the observed GW. For an arbitrarily inclined trajectory, it
suffices to put back the relevant factor $\cos^2\theta$ in the final
results.  Under these simplifying assumptions, the quadrupole moment
of Eq.~\eqref{shapiro2} takes the form
\begin{equation}
\Icut_{k\ell} = m R_\star^2 \left(\bar{x}_k \bar{x}_\ell - \frac{1}{3}
\delta_{k\ell} s^2 \right) \doteq m R_\star^2 \bar{\Icut}_{k\ell}.
\label{pbh1}
\end{equation}
Finally, using the dimensionless time variable $\tau$, the strain
tensor of Eq.~\eqref{shapiro1} becomes
\begin{equation}
h_{ij}^\textsc{tt}
= \frac{2G^2 m M_\star }{c^4 R_\star d}\, \perp_{ijk\ell} 
\bar{\Icut}''_{k\ell}
\doteq h_+ \varepsilon_{ij}^+ + h_\times \varepsilon_{ij}^\times,
\label{pbh2}
\end{equation}
in which the polarization tensors have nonvanishing components only in
the $y-z$ directions, reading $\varepsilon^+ = \sigma_z$ and
$\varepsilon^\times = \sigma_x$ in terms of the Pauli matrices.
Notice that the prefactor in Eq.~(\ref{pbh2}) is a number that
characterizes the physical properties of the system, including its
distance from the observer, while the geometric part
$\bar{\Icut}''_{ij}$ also includes the time dependence. Explicitly,
the modes are given by
\begin{align}
h_+ = & \frac{2G^2 m M_\star }{c^4 R_\star d}\, \qty{ \alpha(\tau) \cos\qty[2\varphi(\tau)]-\beta(\tau)
\sin\qty[2\varphi(\tau)]}
\nonumber \\
\doteq & \left(\frac{2G^2 m M_\star }{c^4 R_\star d}\right)\bar h_+\,, 
\label{scaled+}
\\ h_\times = & 
\frac{2G^2 m M_\star }{c^4 R_\star d}\, \left\{ \alpha(\tau)
\sin[2\varphi(\tau)]+\beta(\tau) \cos[2\varphi(\tau)]
\right\}
\nonumber \\
\doteq & \left(\frac{2G^2 m M_\star }{c^4 R_\star d}\right) \bar h_\times \, ,
\label{scaledx}
\end{align}
where we introduced the notation for the scaled strains $\bar h_+$ and $\bar h_\times$, and defined
\begin{eqnarray}
&\alpha (\tau)& =  
  s'^2 + ss'' - 2s^2\varphi'^2 = s'^2-\frac{\bar \ell^2}{s^2} -\frac{\bar M(s)}{s},
\\
&\beta (\tau) &= s^2 \left(\varphi'' +
4 \frac{s'}{s} \varphi' \right) = 2  \frac{\bar\ell s'}{s},
\end{eqnarray}
so that the resulting $\ell =2$ multipole amplitude reads
\begin{equation}
h^\textsc{tt} (\tau) \doteq \sqrt{h_+^2 + h_\times^2} = \frac{2G^2 m M_\star }{c^4 R_\star d}\, \sqrt{ \alpha^2 + \beta^2}.
\end{equation}

\begin{table}[h]
\centering
\begin{tabular}{|c|c|c|c|}
\hline
$\ \ $ & distance ($d$/km) & Mass ($M_\star$/kg) & Radius ($ R_\star$/km) \\
\hline
Sun & $1.496 \times 10^{8}$  & $M_\odot$ & $6.957\times 10^5$  \\ 
\hline 
Vela & 
$9.072 \times 10^{15}$ & $  1.4 M_\odot$ & 9.656 \\ 
\hline 
\end{tabular} 

\caption{Sizes and distances from Earth of our Sun and the Vela
  pulsar.}

\label{tab:SunVela}
\end{table}
For the sake of comparison and future reference, we consider a PBH of
mass $m$ orbiting either our Sun or the Vela pulsar, with the relevant
physical parameters gathered in Table~\ref{tab:SunVela}. Substituting
these quantities into Eq.~(\ref{pbh2}), it is evident that for PBHs of
identical masses, the ratio of the GW amplitudes emitted by the
Sun-PBH system compared to the Vela-PBH system, as observed on Earth,
is approximately $600$. Hence, the same orbit of a PBH around the Sun
would emit GW radiation that, when measured on Earth, would be
approximately 600 times more intense than the radiation originating
from Vela. This large ratio is interesting given the prevalence of
Sun-like stars within our astronomical neighborhood.


We remark that if two stars have the same mass distribution, 
then PBH orbits with the same (dimensionless) energy $\bar{E}_\textsc{t}$ and
angular momentum $\bar{\ell}$ generate exactly the same pattern of
gravitational radiation from the term $\bar{\Icut}''_{ij}$ in
Eq.~\eqref{pbh2}. Therefore the amplitude given by the coefficient in
Eq.~\eqref{pbh2} provides the most important piece of information
concerning the intensity of the emitted radiation.

Lastly, we note that the energy lost by the system due to the emission
of gravitational waves (within the same quadrupole approximation) is
given by \cite{1986bhwd.book.....S,Maggiore:2007ulw}
\begin{equation}
\frac{\dd E}{\dd t} = -\sum_{ij} \frac{ G}{5 c^5} \left( \frac{\dd^3
  \!\!\Icut_{ij} }{\dd t^3} \right)^2 .
\label{dEdt1}
\end{equation}
As we will see, for the systems of interest here, the gravitational
radiation spectrum is strongly peaked at a frequency $f_\text{peak}
\simeq v_0 /  R_\star$, with $v_0 \lesssim v_\text{esc} = \sqrt{ 2 G M_\star
  /  R_\star}$, or $f_\text{peak}^2 \simeq 2G M_\star / R_\star^3$. We may then
approximate $\dot{h}^\textsc{tt}_{ij} \simeq i (2 \pi f_\text{peak} )
h_{ij}^\textsc{tt}$, and, upon using Eqs.~(\ref{shapiro1}) and
(\ref{dEdt1}),
\begin{equation}
\bigg\vert \frac{\dd E}{\dd t} \bigg\vert \simeq \frac{ 4 \pi^2}{5}
\frac{ c^3 d^2 M_\star}{R_\star^3} \big\vert h^\textsc{tt}\big\vert^2 .
\label{dEdt2}
\end{equation}
The time-scale $t_\text{gw}$ over which energy loss due to emitted
gravitational radiation would back-react on the PBH orbit is set by
$E_\text{grav} / t_\text{gw} \simeq \vert \dd E / \dd t \vert$, where
$E_\text{grav} = G m M_\star /  R_\star$, which yields
\begin{equation}
t_\text{gw} \simeq \frac{ 5}{4 \pi^2} \frac{ Gm R_\star^2}{c^3 d^2}
\frac{1}{\vert h^\textsc{tt} \vert^2} .
\label{tgw1}
\end{equation}
For fiducial values ($m = 10^{20} \, \text{kg}, M_\star = M_\odot, R_\star =
R_\odot, d = 1.5 \times 10^{8} \, \text{km} = 1 \, \text{AU}$), and
using our estimate for $h^\textsc{tt}$ below, in
Eq.~(\ref{hTTtypical}), we then find $t_\text{gw} \sim 10^{24} \,
\text{s}$ for cases of interest here. Given $t_\text{orbit} \sim 10^4
\, \text{s}$ and $t_\text{dyn} \sim 10^{13} \, \text{s}$ for these
same fiducial parameters, we thus confirm the strict hierarchy
\begin{equation}
t_\text{orbit} \ll t_\text{dyn} \ll t_\text{gw}
\label{thierarchy}
\end{equation}
for the PBH-star systems we are interested in. Hence we will neglect
both dynamical friction and energy loss from gravitational radiation
in what follows.

\section{Simulations}
\label{simulations}

\begin{figure}[h]
\includegraphics[width=\linewidth]{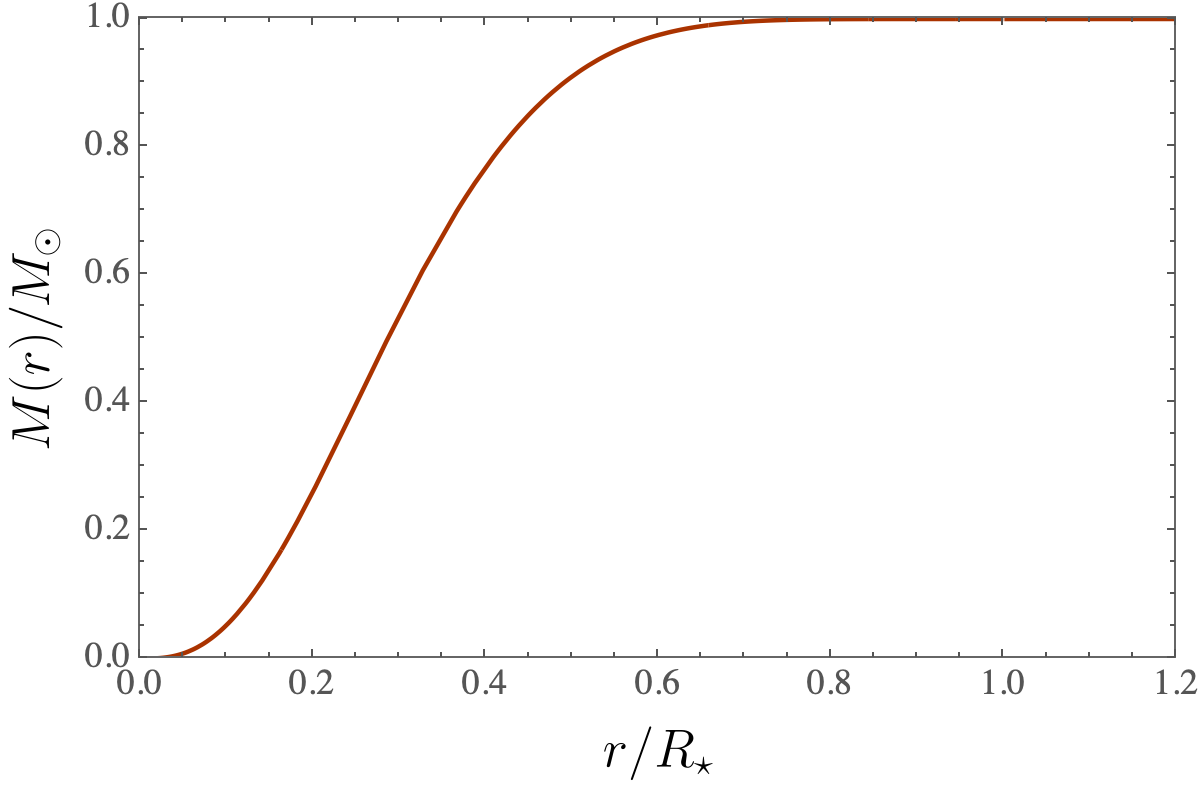}

\caption{The normalized mass of the star, as determined from the
  idealized model given by $\rho(r)$ in Eq.~(\ref{rho}).}

\label{mass}
\end{figure}

Let us now investigate the GW pattern emitted by a PBH in a bound
orbit around a typical star, like our Sun. For the sake of simplicity,
we will adopt an idealized model \cite{2024arXiv240508113D} describing
the mass-density profile\footnote{It is interesting to note that if a constant density profile is
assumed, the corresponding gravitational potential experienced by the
orbiting particle becomes harmonic ($\sim r^2$) when the particle is
in the interior of the star and Keplerian ($\sim r^{-1}$) when it is
in the exterior region. Thus, according to Bertrand's theorem
\cite{2002clme.book.....G}, all bounded solutions will result in
closed orbits if the particle's path is entirely inside the star or
entirely outside it. However, for hybrid orbits, the trajectories will generally be open but could be closed for specific initial conditions \cite{2024arXiv240508113D}.} of the star $\rho(r)$ given by
\begin{equation}
    \rho(r) = \rho(0) \left(1-
    \frac{r}{ R_\star}\right)^6 \Theta \left(1 - \frac{r}{ R_\star}\right),
    \label{rho}
\end{equation}
where $\rho(0) = 1.184 \times 10^5\, \text{kg}\, \text{m}^{-3}$, and
$\Theta(x)$ is the Heaviside step function, defined as $\Theta(x) = 1$
when $x>0$, $\Theta(0) = \frac12$, and $\Theta(x) = 0$ when $x<0$. The
value for $\rho(0)$ is then chosen such that the total mass of the
star coincides with the Sun's mass. For the purposes of our
discussion, this simple model describes the behavior of typical stars
sufficiently well. (See Ref.~\cite{2024arXiv240508113D} for further details.) The mass
function $M(r)$ introduced in the last section can now be directly
obtained by integrating $\rho(r)$ and is depicted in Fig.~\ref{mass}.

\begin{figure}[h]
\includegraphics[width=\linewidth]{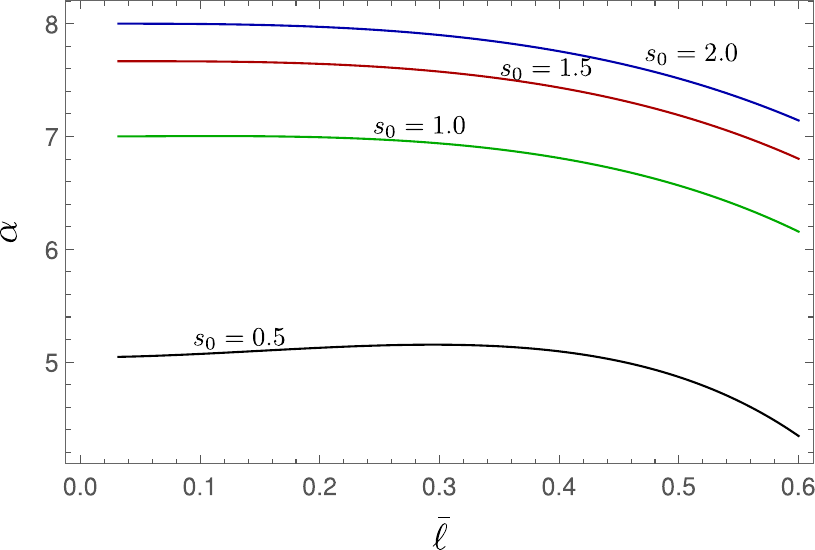}

\caption{Maximum amplitude of the gravitational wave signals emitted
  by the system, $h_+$ and $h_\times$, as a function of $\bar\ell$,
  for some representative values of $s_0$.}
\label{maxhfigure}
\end{figure}

The graph displayed in Fig.~\ref{maxhfigure} is constructed by
finding, for each \(0 < \bar{\ell} < 1\), the smallest root of the equation
\begin{equation}
\bar{V}(s) = \frac{\bar{\ell}^2}{2 s^2} + \int_{+\infty}^{s}
\frac{\bar{M}(u)}{u^2} \dd u = \bar{V}(s_0),
\end{equation}
i.e., the minimum distance of the orbit that starts at $s = s_0$ with
zero radial velocity. The value of scaled strain \(\bar h_+\) at this point is then
computed, assuming it is aligned with the \(x\)-axis. This value
coincides with the maximum values of \(\bar h_+\) and \(\bar h_\times\)
across all orbits with the same fixed angular momentum and energies
within the range \(\bar{V}_\text{min} \leq \bar{E}_\textsc{t} \leq\bar{V}(s_0)\).
This maximum value of $\bar h_+$ is obtained by means of 
\begin{equation}
\alpha_\text{max} = \frac{\bar{\ell}^2}{s_\text{min}^2} +
\frac{\bar{M}(s_\text{min})}{s_\text{min}},
\label{almax}
\end{equation}
where $s_\text{min}$ is taken as the minimum distance of the orbit to the center of the star, as explained above, and is also a function of $\bar \ell$.

\begin{figure}[h]
\includegraphics[width=\linewidth]{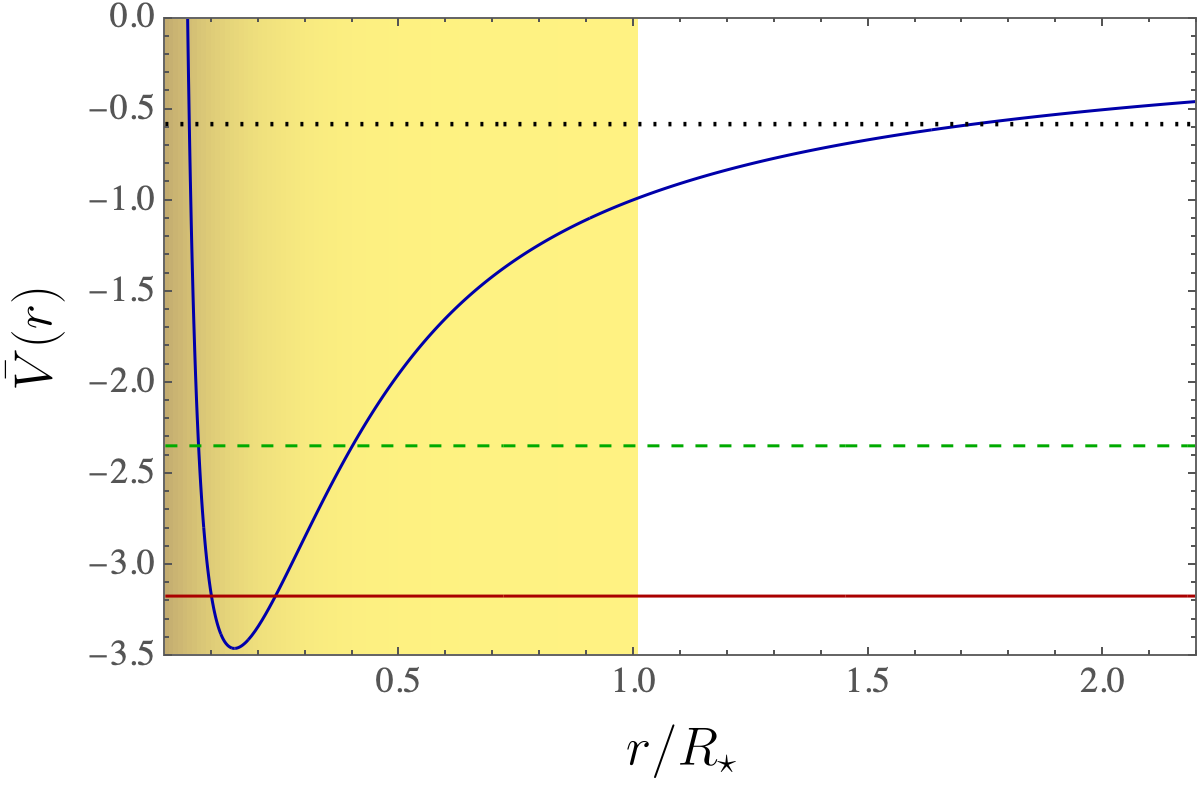}

\caption{The effective potential energy of the PBH as a function of
  $s=r/ R_\star$, where $ R_\star$ is the radius of the star. Here we assumed an
  angular momentum per unit mass such that $\bar\ell = 0.1410$. The
  three horizontal lines indicate orbits with three distinct values of
  the total energy. The dotted horizontal line (with $\bar
  E_\textsc{t} > -1$ ) corresponds to a hybrid orbit, while the other
  two horizontal lines are associated with inner orbits.}

\label{fig1}
\end{figure}

In Fig.~\ref{fig1} the potential energy of the PBH is shown as
a function of the distance $r$ for selected initial
conditions. In particular, the dimensionless angular momentum per unit
mass was chosen to be $\bar\ell = 0.141$, from which, using the data for
the Sun ($M_\odot = 1.989\times 10^{30} \,\text{kg}$), it follows that
$\ell = 4.28\times 10^{13}\,\text{m}^2\text{s}^{-1}$. The shaded
gradient region represents the density of the mass profile function
described by Eq.~(\ref{rho}).

Three possible values for the energy of the particle are also
represented (the horizontal lines), corresponding to orbits with
different initial conditions. The solid line near the bottom of the
effective potential is associated with an orbit such that the total
energy of the particle is given by $\bar E_\textsc{t} =-3.1678$
($E_\textsc{t} =-6.0452\times 10^{31} \,\text{J}$ for a PBH of
$10^{20} \,\text{kg}$).
The dashed line corresponds to an orbit that achieves the maximum and
minimum distances from the center of attraction at $r=0.39856 \, R_\star$ and
$r=0.070740 \,  R_\star$, respectively, where the corresponding speeds are
approximately $v\approx 154.51 \, \text{km}\,\text{s}^{-1}$ and
$v\approx 870.54\, \text{km}\,\text{s}^{-1}$.
Higher than $\bar E_\textsc{t} = -1$ values of energy would lead to
orbit solutions that advance to the exterior region ($r> R_\star$), as the
one depicted by the dotted straight line, that achieves a maximum
distance from the center at $r=1.7229\, R_\star$. In this solution the particle
speed at perihelion is $v \approx 357.43 \,\text{km}\,\text{s}^{-1}$
while at the aphelion it achieves $v\approx 1,208.7
\,\text{km}\,\text{s}^{-1}$.

\begin{figure}[t]
\centering
\includegraphics[width=\linewidth]{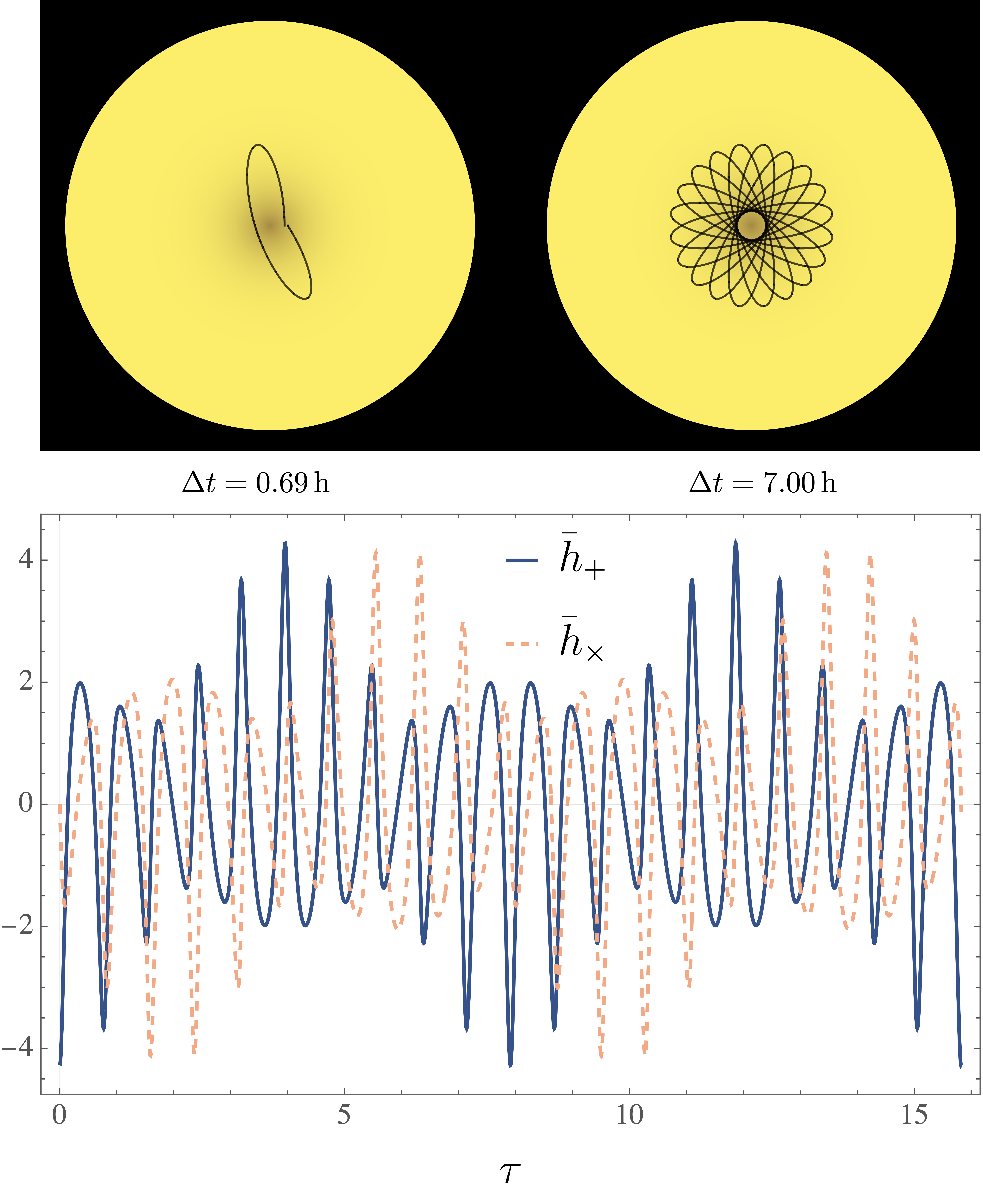}

\caption{The top panels show the orbit of a PBH of total energy given
  by $\bar E_\textsc{t} =-2.3435$ and with an angular momentum such
  that $\bar\ell = 0.141$. The top-left panel illustrates the path of
  the PBH in a complete revolution $\Delta\varphi=2\pi$, while the
  top-right panel shows a complete closed orbit. The corresponding GW scaled strains
  $\bar h_+$ and $\bar h_\times$ emitted by the system when performing the closed orbit are shown in the bottom panel, as a function of
  dimensionless time $\tau$.}
\label{fig:Case1}
\end{figure}

\begin{figure}[t]
\centering
\includegraphics[width=\linewidth]{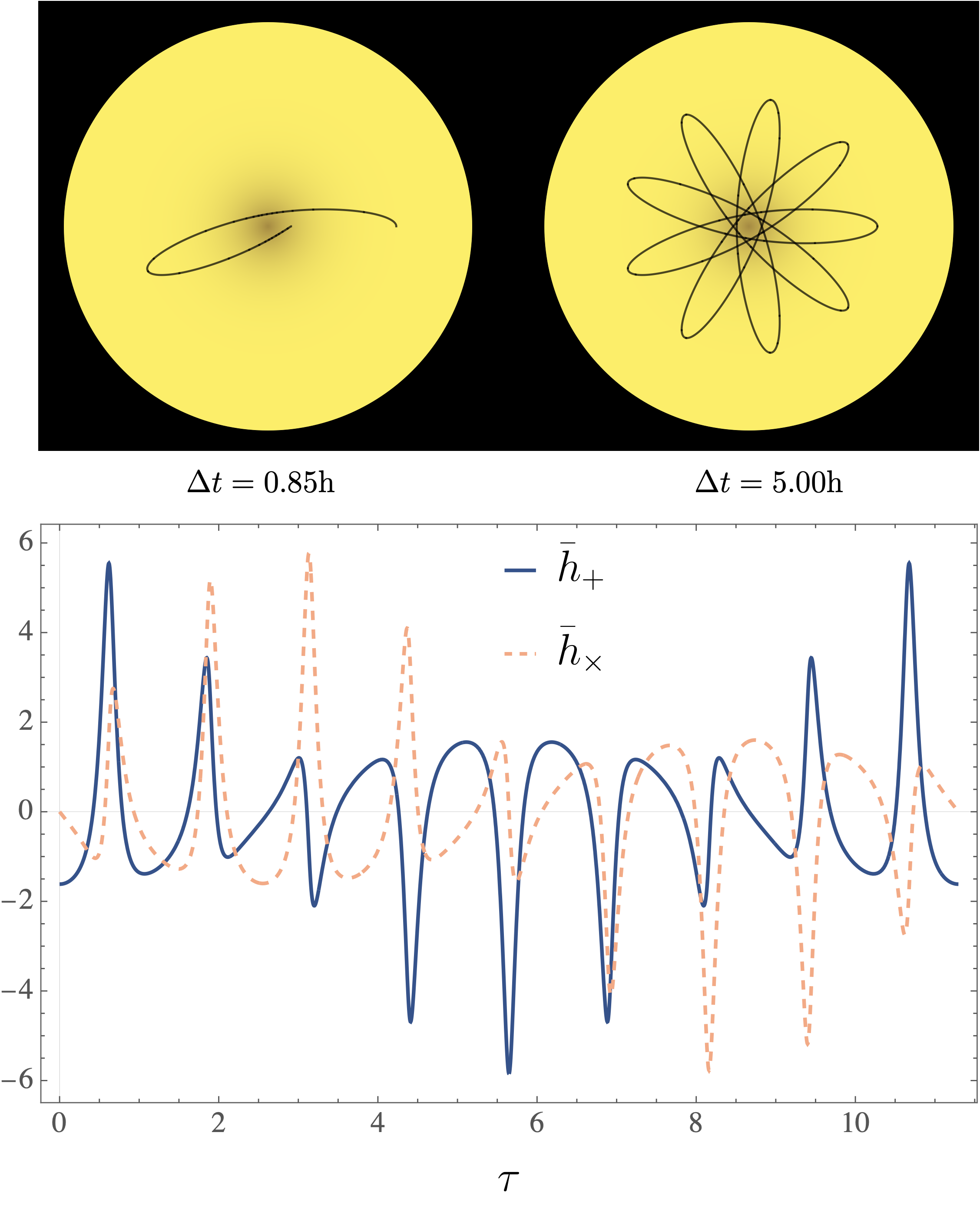}

\caption{The top panel shows the orbit of a PBH of total energy given
  by $\bar E_\textsc{t} = -1.56456$ and with an angular momentum such
  that $\bar\ell = 0.141$. The corresponding GW scaled strains $\bar h_+$ and $\bar h_\times$ emitted by the system during a complete closed orbit are
  shown in the bottom panel.}
\label{fig:Case2}
\end{figure}

\begin{figure}[t]
\centering
\includegraphics[width=\linewidth]{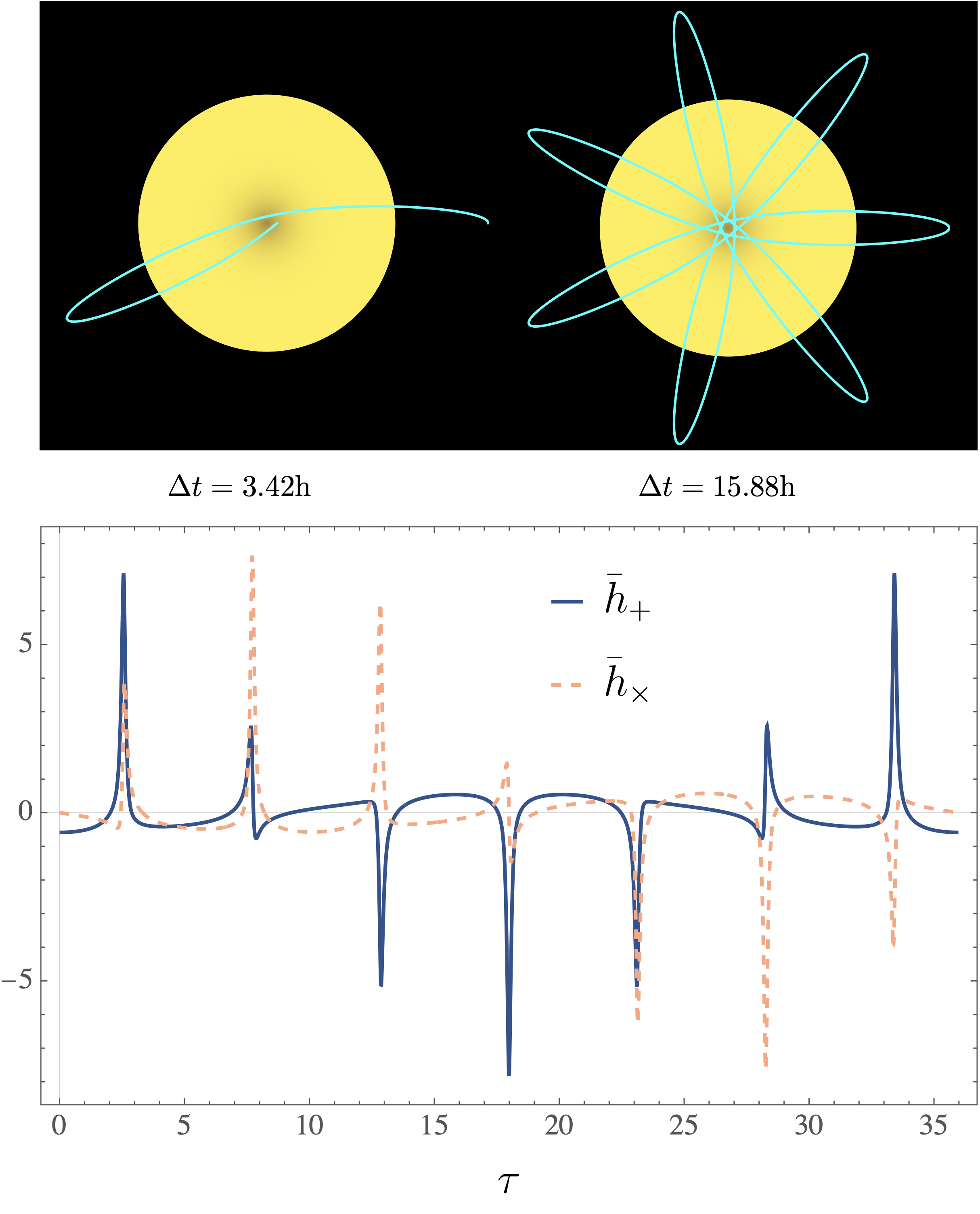}

\caption{The top panel shows a closed semi-interior orbit of a PBH of
  total energy given by $\bar E_\textsc{t} = -0.57707$ and with an
  angular momentum such that $\bar\ell = 0.141$. The bottom panel
  shows the corresponding GW scaled strains $\bar h_+$ and $\bar h_\times$ emitted by
  the system when performing one closed orbit, as a function of
  dimensionless time $\tau$.}
\label{fig:Case3}
\end{figure}

The trajectories for the three solutions discussed in Fig.~\ref{fig1}
are shown in the top panels of Figs.~\ref{fig:Case1}, \ref{fig:Case2}
and \ref{fig:Case3}. Notice that the least eccentric orbit corresponds
to the solution with the smallest total energy, as could have been
anticipated by examining Fig.~\ref{fig1}. The initial conditions were
chosen in such a way to produce closed orbits~\cite{2024arXiv240508113D}.

The GW signals are depicted in the bottom panels of these figures. For
instance, in Fig.~\ref{fig:Case1}, the top-left panel illustrates the
path of a PBH covering a full $2\pi$ angular span, which takes
approximately $0.69\,\text{h}$ to complete. The top-right panel shows
the entire closed orbit, and the corresponding map of the emitted GW
is shown in the bottom. Note that the interval of time for a closed
orbit is about $7\,\text{h}$. As it is periodic, the complete signal
repeats every $7\,\text{h}$, leading to a frequency of $3.97\times
10^{-5}\, \text{Hz}$. However, the interval between two successive
maxima of amplitude is about $0.35\,\text{h}$, which leads to a
frequency of about $8.05\times 10^{-4} \,\text{Hz}$. The same
reasoning applies to Figs.~\ref{fig:Case2} and \ref{fig:Case3}, where
the amplitude of the signal becomes larger as the eccentricity of the
orbit increases.

In particular, Fig.~\ref{fig:Case3} illustrates a closed semi-interior
orbit with an initial condition of $r(0)=2 R_\star$, corresponding to the
maximum distance the PBH reaches along its path. As can be seen, the
strain signals are more intense and sharper than in the other less
eccentric orbits. In this specific case, the amplitude of the strains
almost achieves the maximum value predicted in Fig.~\ref{maxhfigure}.

It can be inferred from the above figures that the strain signals
become sharper and more pronounced as the eccentricity of the orbits
increases. This fact suggests that an orbit with null angular momentum
(a free falling radial orbit) would maximize the amplitude of the
strains.  Suppose a PBH falls radially toward the star, starting at
rest at $s_0$. At this point, its total energy is given by 
$\bar{E}_\textsc{t} = \bar{V}(s_0)$, where Eq.~\eqref{vbar} takes the
form
\begin{equation}
\bar{V}(s) = 
\begin{cases} 
-\displaystyle \int_s^1 \frac{\bar{M}(u)}{u^2} \dd u - 1, & \text{if } s < 1, \\
& \\
-1/s, & \text{if } s > 1.
\end{cases}
\end{equation}
As the PBH passes through the center of the star, energy conservation
requires that $\frac{1}{2}s'(\tau)^2 + \bar{V}(0) = \bar{V}(s_0)$.
However, $s_0 > 1$ for a semi-interior orbit, and therefore as the PBH
passes through the center, its velocity is such that $s'^2 =
2\left[-1/s_0 - \bar{V}(0)\right]$.
Furthermore, the scaled strain $\bar h_+$ associated with this type of orbit has
the form $\bar h_+ = s'^2 -  \bar{M}(s)/s$, which reaches a maximum amplitude when $s = 0$. In this case,
\begin{align}
\bar h_+ = s'^2 = -2 \left[\frac{1}{s_0} + \bar{V}(0)\right],
\end{align}
which achieves its largest value when $s_0 \to \infty$, such that
\begin{equation}
\lim_{s_0 \to \infty} \bar h_+ = - 2\bar{V}(0) = 2\left[1 + \displaystyle\int_0^1\frac{\bar{M}(u)}{u^2} \dd u \right].
\end{equation}
Additionally, note that this value depends on the full mass
distribution. For the particular mass-density distribution defined by
Eq.~\eqref{rho}, we get $\bar{V}(0) = -9/2$ and hence the maximum 
value achieved by this strain is such that $\bar h_+ = 9-2/s_0$. 
If we choose $s_0=1.7$, as in Fig.~\ref{fig:Case3}, the
maximum strain is $\max (h_+) \approx 7.8$, which is close to the
limit of the plot in Fig.~\ref{maxhfigure} as $\bar{\ell} \rightarrow
0$. Note that the maximum possible amplitude is $9$, which corresponds to
a radially free-fall trajectory starting from infinity. We
emphasize that this maximum value depends on the star's mass
distribution.

In order to have an estimate of the effect that would be measured in a
GW detector, suppose a PBH of mass $m$ is orbiting the Sun, as
described in any of the solutions depicted in the above figures. The
maximum amplitude $h^\textsc{tt}$ of the gravitational wave emitted by
this system that would be received on Earth would be:
\begin{equation}
\begin{split}
h^\textsc{tt} \approx & \;2.11\times 10^{-24} \alpha_\text{max}
\left(\frac{m}{10^{20}\,\text{kg}}\right)
\left(\frac{M_\odot}{1.99\times 10^{30} \,\text{kg}}\right) \nonumber
\\ &\times \left(\frac{1.50\times
  10^{11}\,\text{m}}{d}\right)\left(\frac{6.96\times
  10^{8}\,\text{m}}{ R_\odot}\right),
  \end{split}
\label{hTTtypical}
\end{equation}
where $\alpha_\text{max}$, given by Eq.~\eqref{almax},
denotes the amplitude of the GW signal.
For example, for a PBH with mass $m = 10^{20} \,
\text{kg}$ orbiting the Sun, the maximum amplitude of the signal
received on Earth would be $h^\textsc{tt} \approx 10^{-23}$. 
We may compare such signals with those predicted from a more strongly relativistic system. Ref.~\cite{2024arXiv240201838B} considers the GW spectrum from a bound PBH undergoing an interior orbit within a neutron star (NS), with $m = 1.4 \times 10^{-6} \, M_\odot \simeq 2.8 \times 10^{24} \, \text{kg}$ and $M = 1.4 \, M_\odot = 2.8 \times 10^{30} \, \text{kg}$ at a distance $d = 10 \, {\rm kpc}$ from Earth. They find typical GW strains $h^{\textsc{tt}} \sim {\cal O} (10^{-24})$ \cite{2024arXiv240201838B}. We may extrapolate our own results to a system involving the same PBH mass orbiting within our own Sun, which yields  
$h^\textsc{tt} \approx 3 \times 10^{-20} \alpha_\text{max} \sim {\cal O} (10^{-19})$. This is significantly stronger than the signal at Earth expected from a typical NS-PBH system.

\section{Detecting GW Signals from Individual Systems}
\label{sec:GWlocalsource}

In this section we consider the possibility of detecting GW signals
from a single PBH of mass $m$ orbiting a Sun-like star, whose mass and
radius we take to be $M_\odot$ and $R_\odot$, respectively. As we will
see, if the PBH-star system is relatively close to the Earth---that
is, if it is bound within the Milky Way galaxy rather than undergoing
Hubble flow---then the typical GW signals would achieve maximum
amplitude for observed frequencies at a near-Earth detector of order
$f \sim {\cal O} (10^{-3} \, \text{Hz})$. Such signals would be
interesting candidates for detection by the \textsmaller{LISA} gravitational-wave
observatory
\cite{Maggiore:2007ulw,LISA:2022yao,Robson:2018ifk,Smith:2019wny}. (Given the typical frequencies expected from such PBH-star systems and the peak sensitivities expected for other upcoming GW detectors, such as the Einstein Telescope and Cosmic Explorer---each of which will be optimized for GW signals with $f \sim {\cal O} (10^0 - 10^3 \, {\rm Hz})$ \cite{Reitze:2019iox,Abac:2025saz}---we do not expect the GW sources considered here to be candidates for detection with those other experiments.)

To begin, we consider a source that produces time-series waveforms
$h_+ (\tau)$ and $h_\times (\tau)$, akin to those calculated from
Eq.~(\ref{pbh2}) and shown for various initial conditions in
Figs.~\ref{fig:Case1}, \ref{fig:Case2}, and \ref{fig:Case3}. After
converting from dimensionless time $\tau$ to source-frame time $t$ (in
seconds), we sample the time-series data at a frequency of 2 Hz, as
appropriate for \textsmaller{LISA} sensitivity up to $\sim 1$ Hz. We then implement
the algorithm of Ref.~\cite{cooley1965algorithm} to compute the
discrete Fourier transforms of the sampled time-series data to yield
$\tilde{h}_+ (f)$ and $\tilde{h}_\times (f)$.

Because the waveforms $h_{+,\times} (\tau)$ are dominated by the
quadrupole moment, we may express the frequency-domain strains as
\cite{Robson:2018ifk,Smith:2019wny,Maggiore:2007ulw}
\begin{equation}
    \begin{split}
        \tilde{h}_+ (f) &= A (f) \left( \frac{ 1 + \cos^2 \eta}{2}
        \right) e^{i \Psi (f)} , \\ \tilde{h}_\times (f) &= i A (f) \,
        \cos \eta \, e^{i \Psi (f)} ,
    \end{split}
\label{tildehAPsi}
\end{equation}
where $A (f)$ is the amplitude, $\Psi (f)$ is the phase, and $\eta$ is
the inclination of the plane of the PBH orbit with respect to the
observer. Given the amplitude $A(f)$, we may construct the signal
power spectral density averaged over sky locations, GW polarizations,
and inclination angles $\eta$ \cite{Robson:2018ifk},
\begin{equation}
S_h (f) = \frac{ A^2 (f)}{2 T_\text{sam}} ,
\label{Shdef}
\end{equation}
where $T_\text{sam}$ is the sampling interval, which we take to be the
period of one complete PBH orbit. For long-lived, continuous-wave GW
signals like the ones we consider here, an instrument like \textsmaller{LISA} can
boost signal-to-noise compared to the instantaneous strain by using
template-matching signal extraction. Taking this into account, we may
calculate the angle-averaged square of the optimal signal-to-noise
ratio \cite{Robson:2018ifk},
\begin{equation}
\langle\rho^2\rangle = \frac{16}{5} \int_0^\infty \frac{ 2 f
  T_\text{obs} S_h (f)}{S_n (f)} \frac{\dd f}{f},
\label{rhodef}
\end{equation}
where $S_n (f)$ is the power spectral density of noise in the \textsmaller{LISA}
detector averaged over angle and waveform polarization, and
$T_\text{obs}$ is the duration of observation. The factor
$(2fT_\text{obs})$ in the numerator arises from improved signal
discrimination via template-matching. For the \textsmaller{LISA} detector, we use
the parameterization of $S_n (f)$ in Ref.~\cite{Robson:2018ifk}.

\begin{figure}[h!]
\centering
\includegraphics[width=0.43\textwidth]{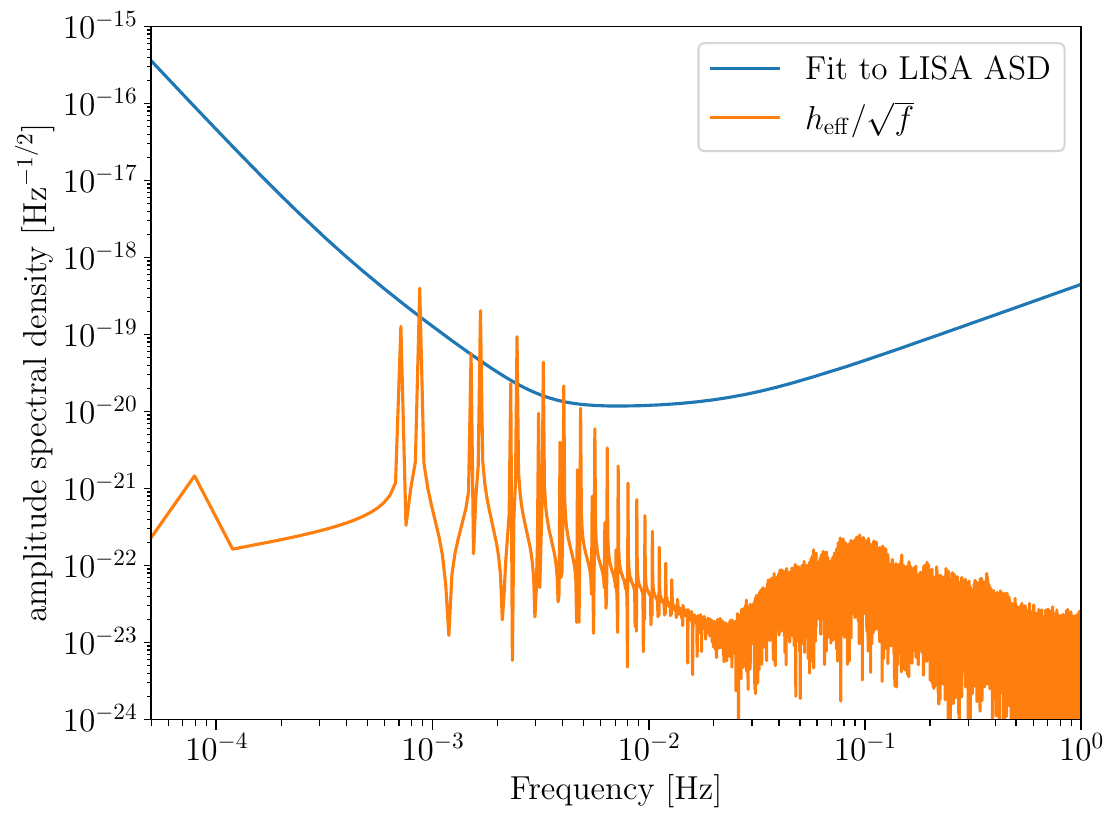} \\
\includegraphics[width=0.43\textwidth]{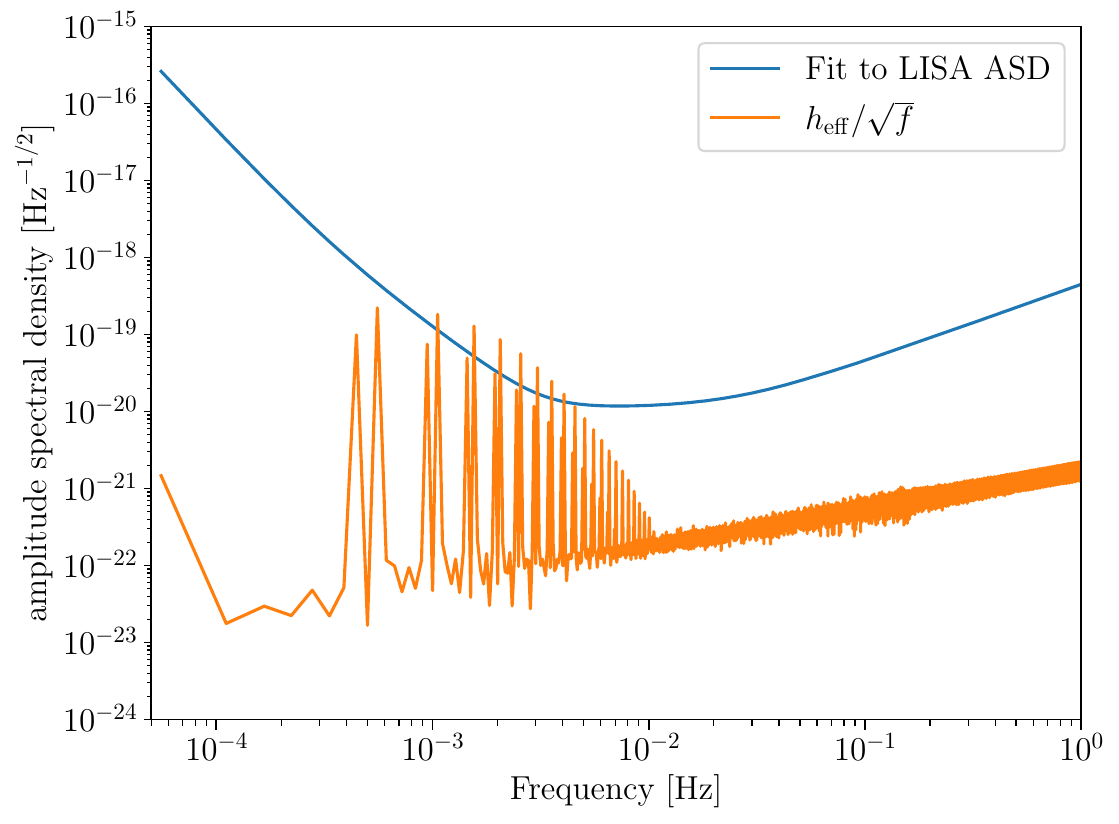}
\includegraphics[width=0.43\textwidth]{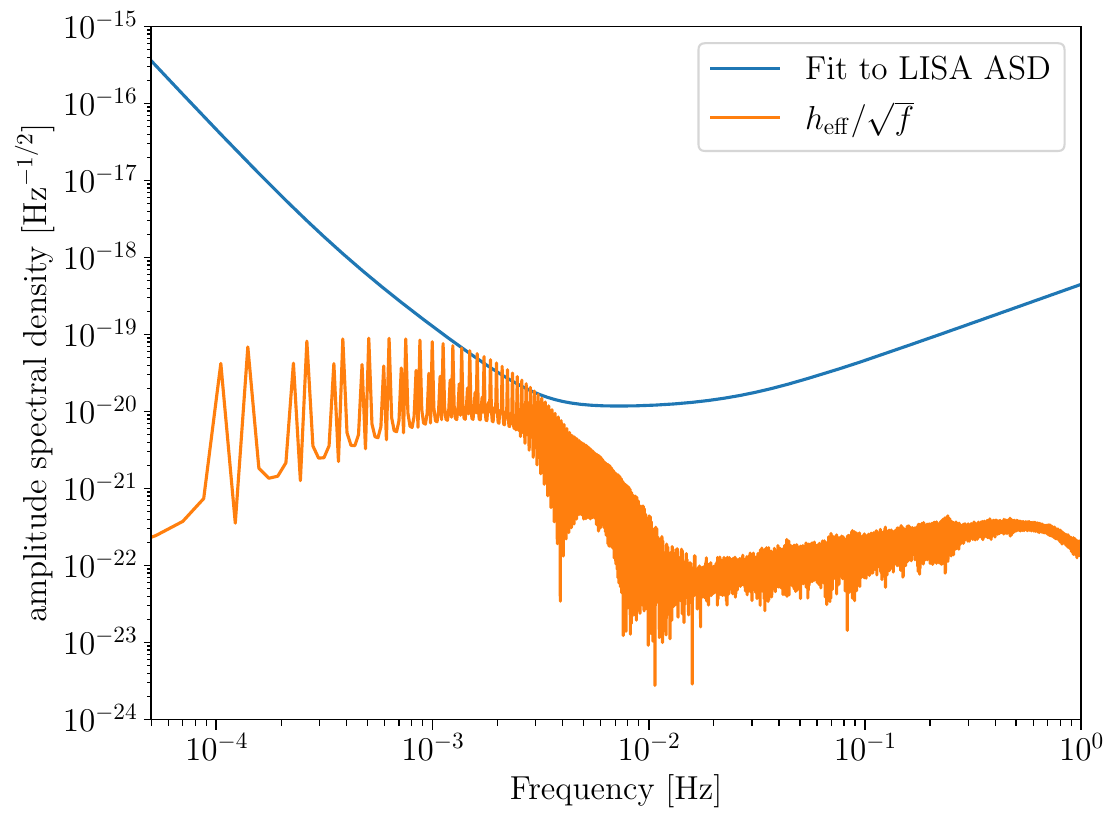}

\caption{\textsmaller{LISA} amplitude spectral density (ASD, in blue) compared to
  $h_\text{eff}$ (orange) for various PBH-star orbital
  configurations. In each case we use $M_\star = M_\odot$ and $R = R_\odot$
  for the host star, $m = 10^{-9} \, M_\odot$ for the PBH mass, and
  set $d = 1 \, \text{AU}$. The amplitude for the signal $h_\text{eff}
  (f)$ scales as in Eq.~(\ref{hTTtypical}) for other selections of
  orbital parameters. ({\it Top}) The initial conditions as in
  Fig.~\ref{fig:Case1}, which yields $\rho = 1.1$ over the 4-year \textsmaller{LISA}
  mission lifetime. ({\it Middle}) The initial conditions as in
  Fig.~\ref{fig:Case2}, which yields $\rho = 1.0$ over the 4-year \textsmaller{LISA}
  mission lifetime. ({\it Bottom}) The initial conditions as in
  Fig.~\ref{fig:Case3}, which yields $\rho = 0.6$ over the 4-year
  \textsmaller{LISA} mission lifetime. In each plot, we have averaged over sky
  location, GW polarizations, and orientation of the PBH orbit with
  respect to the detector.  }

\label{fig:LISAsignals}
\end{figure}

The signal-to-noise ratio $\rho$ is computed by integrating over all
frequencies. To compare the effective signal-to-noise at a given
frequency $f$, we may define the quantity \cite{Robson:2018ifk}
\begin{equation}
    h_\text{eff}^2 (f) \equiv \frac{16}{5} f \left( 2 f T_\text{obs}
    \right) S_h (f) .
\label{heffdef}
\end{equation}
In Fig.~\ref{fig:LISAsignals}, we plot $h_\text{eff} (f)$ and the
amplitude spectral density $\sqrt{S_n (f)}$ for \textsmaller{LISA} for the GW
waveforms shown in Figs.~\ref{fig:Case1}, \ref{fig:Case2}, and
\ref{fig:Case3}. In each case, we have set $m = 2 \times 10^{21} \,
\text{kg} = 10^{-9} \, M_\odot$ and $d = 1.5 \times 10^{8} \,
\text{km} = 1 \, \text{AU}$. For these parameters, we see that the
typical $h_\text{eff} / \sqrt{f} \sim \sqrt{S_n}$ across the \textsmaller{LISA} sensitivity
range. Over the full \textsmaller{LISA} mission, with $T_\text{obs} = 4 \,
\text{yr}$, we find $\rho \gtrsim {\cal O} (1)$ for several
configurations. Given the form of Eq.~(\ref{pbh2}), the amplitude
$A(f)$, and therefore the signal-to-noise ratio $\rho$, scales
linearly with $m$ and $M_\star$ and inversely with $R_\star$ and $d$.

As expected, each of the GW waveforms shown in
Fig.~\ref{fig:LISAsignals} peaks at a frequency $f \simeq 1 /
t_\text{orbit}$. For the cases shown there, with $v_0 \sim 10^2 \,
\text{km} \, \text{s}^{-1}$ and semi-major axis of the PBH orbit
${\cal R}_\text{pbh} \simeq R_\odot$, we find $f_\text{peak} \simeq
10^{-3} \, \text{Hz}$. In fact, upon plotting the waveforms
$\tilde{h}_{+,\times} (f)$ rather than the weighted combination
$h_\text{eff} (f)$, we find $A_\text{peak} \equiv A (f_\text{peak})
\sim 10^2 A (f_\text{next})$, where $A (f_\text{next})$ is the
amplitude of the next-leading Fourier component.

For the curves shown in Fig.~\ref{fig:LISAsignals}, we have averaged
over sky locations when evaluating $\rho$. The results reveal an interesting trade-off: although a larger initial orbital distance enhances the  amplitude of the emitted GW signal, it simultaneously leads to a lower signal-to-noise ratio, reducing the detectability of the event. In contrast, more confined orbits produce GW signals with a lower amplitude that can be detected more easily, since $\rho \geq 1$.
As discussed in Refs.~\cite{Robson:2018ifk,Smith:2019wny}, if one knows the location
of a given source, then the optimal value of $\rho$ can be improved by
not performing an average over the full sky. In the present case, we
remain agnostic as to where a given source might appear and hence we
perform the typical all-sky averaging.

For $m = 10^{-9} \, M_\odot$ and $M_\star = M_\odot$, the orbits as
simulated here should remain unaffected by dynamical friction up to a
time-scale $t_\text{dyn} \sim 10^{-1} (M_\odot / m) \, t_\text{orbit}
\sim 10^5 \, \text{yr}$. On the other hand, if a signal detectable by
\textsmaller{LISA} were to come from a small-mass PBH orbiting a Sun-like star other
than our own Sun --- and hence at a larger distance from the Earth
than $d = 1 \, \text{AU}$ --- then the PBH mass $m$ would need to be
correspondingly larger than $10^{-9} \, M_\odot$. As a concrete
example, for a PBH in orbit around the star Proxima Centuri, at a
distance $d = 4.0 \times 10^{13} \, \text{km} = 2.7 \times 10^5 \,
\text{AU}$ from the Earth, the PBH mass $m$ would need to be $m = 2.7
\times 10^{-4} \, M_\odot$ in order to yield $\rho > 1$ for a \textsmaller{LISA}
detection. With that mass, $t_\text{dyn} \sim 1 \, \text{yr}$, making
the likelihood for such a detection with the \textsmaller{LISA} detector quite
small.

Lastly, we note that for the plots in Fig.~\ref{fig:LISAsignals}, we have used the simple quadrupole approximation when evaluating the waveforms $h_{+,\times}$, which should be sufficiently accurate for GWs in the far-field region arising from the systems we consider here. Yet if a PBH were orbiting our own Sun, the typical wavelength of the GWs, $\lambda \sim c / f_{\rm peak}$, would be comparable to the distance between the Sun and the \textsmaller{LISA} detectors, $d \sim 1 \, {\rm AU}$. Our examples show a maximum
peak frequency $f^\text{max}_\text{peak} \sim 5\times 10^{-3}\,\text{Hz}$,
hence a wavelength $\lambda_\text{peak}^\text{min} \sim 0.3 R^\text{max}_\textsc{lisa}$, which is comparable to but slightly shorter than the distance of the source to the observer (in this case, the \textsmaller{LISA} detectors). In this intermediate
zone, one should arguably take into account higher-order terms in
the post-Minkowskian expansion~\cite{Blanchet:2013haa}. We expect that given such additional effects, the estimates presented here would serve as lower limits to the amplitude of the full signals, though how such signals would register in the interferometer remains to be clarified. We leave to future work the interesting question of how near-field radiation effects might alter the predicted signals shown in Fig.~\ref{fig:LISAsignals}. 

When considering GW signals from PBHs within our own Solar System, one must also consider possible background noise from actual asteroids, since we are considering PBHs within the asteroid-mass range.   
Given the estimates above, the amplitude of the GW signal produced by an asteroid would be maximal provided its trajectory is as close as possible to the Sun, therefore leading to a signal of the same order of magnitude and within the same range of frequencies as those considered in this section. However, such a near approach by an actual asteroid (rather than by a PBH) is quite unlikely: an asteroid would probably be destroyed by the Sun by other effects before it got as close to the Sun as $R_\odot$. One might next consider an asteroid that comes within a minimal distance $R_\text{min}$ of the Sun, with $R_\text{min}$ between $\sim 10 \, R_\odot$ and, say, the radius of Mercury's orbit ($\sim 100 \, R_\odot$).\footnote{For comparison, Mercury itself would emit quadrupolar GWs with an amplitude roughly two orders of magnitude smaller than that produced by the systems investigated in Figs.~\ref{fig:Case1}--\ref{fig:Case3}. Note however that such an estimate is not physically meaningful for an observation point that lies well within the near zone of the source. It is provided here only to illustrate the order of magnitude of the effects involved.} The typical frequency range of GW emission would then be redshifted by an amount $R_\text{min}/ R_\odot$, expected to be anywhere between $10$ and $100$, while the amplitude would be reduced by the same amount. Given the \textsmaller{LISA} sensitivity curve, this would immediately render such a signal undetectable. Finally, let us emphasize that all relevant asteroids have hyperbolic trajectories and therefore come close to the Sun only once: even if misleadingly interpreted as coming from a PBH, such a signal would not repeat itself, yielding an easy discrimination from the PBH signals analyzed here.



We further note that for the class of objects considered in this study, the orbital periods are significantly shorter than those of any other known bodies bound to the Sun. For instance, in the cases shown in Figs.~\ref{fig:Case1}--\ref{fig:Case3}, the orbital periods are of the order of one hour, whereas the asteroids with the closest known orbits around the Sun have periods well above 100 days. Therefore, even in situations in which corrections to the quadrupole approximation due to near- or intermediate-zone effects become relevant, the characteristic pattern of an emitted signal remains tied to the orbital period. This pronounced difference in timescales between PBHs and other bodies would allow an eventual \textsmaller{LISA} detection to unambiguously identify a PBH as the source of the observed GWs.

\section{Contribution to the Stochastic GW Background}
\label{sec:SGWB}

If a significant fraction of the dark matter consists of small-mass
PBHs, then such objects must have been ubiquitous throughout cosmic
history. Likewise, Sun-like stars have been common throughout the
universe since around redshift $z = 3$, that is, over the past 11.5
Gyr \cite{DunlopStarFormation}. If the capture rate for small-mass
PBHs by Sun-like stars is not negligible, then a significant
population of bound PBH-star systems should have formed throughout the
universe over time. The gravitational-wave emissions from a population
of such independent sources would contribute to the stochastic GW
background (SGWB). In this section we consider what contribution we
might expect from PBH-star orbits to the SGWB, and whether such a
contribution might be detectable via pulsar timing arrays
\cite{NANOGrav:2023gor,EPTA:2023fyk,Reardon:2023gzh,Xu:2023wog}.

We follow Ref.~\cite{Lehmann:2022vdt} to estimate the PBH capture
rate. (See also
Refs.~\cite{Khriplovich:2009jz,Capela:2012jz,Capela:2013yf,Lehmann:2020yxb,Genolini:2020ejw,Caplan:2023ddo,2024arXiv240408057C,Santarelli:2024uqx,Bhalla:2024jbu}.)
Ref.~\cite{Lehmann:2022vdt} considers several mechanisms that would
yield a bound PBH orbiting a star, including energy loss by the PBH
due to GW emission, dissipative dynamics such as gas drag and
dynamical friction, and three-body capture and ejection, involving
exchange of energy between the PBH, its host star, and a Jupiter-like
planet. As noted above, for PBHs in the mass range of interest here,
energy loss via GW emission remains weak, and dissipative dynamics are
most effective in denser media such as gas clouds undergoing early
star formation, whereas three-body capture can occur at any time over
cosmic history. For the three-body scenario,
Ref.~\cite{Lehmann:2022vdt} estimates an equilibrium number of bound
PBHs per star of the form
\begin{equation}
\begin{split}
N_\text{eq} &\simeq \left[ \frac{0.65 + \log_{10} ( M_\star /
    M_\text{planet} )}{3.7} \right] \left( \frac{
  M_\star}{M_\odot} \frac{ {\cal R}_\text{planet}}{5 \,
  \text{AU}} \right)^{3/2} \\ &\quad \times \left( \frac{ v_0}{220 \,
  \text{km} \, \text{s}^{-1}} \right)^{-3} \left( \frac{
  \rho_\textsc{dm}}{0.4 \, \text{GeV} \, \text{cm}^{-3}} \, \frac{
  10^{14} \, \text{kg}}{m} \right),
\end{split}
\label{Ncapture}
\end{equation}
where ${\cal R}_\text{planet}$ is the semi-major axis of the planet's
orbit around its host star, $v_0$ is the typical PBH velocity in the
vicinity of the stellar system (prior to capture), and
$\rho_\text{DM}$ is the local dark matter energy density. The fiducial values
are selected for the Sun-Jupiter system while also assuming that PBHs
constitute all or most of the dark matter. For PBHs within the
asteroid-mass range, one may therefore expect $N_\text{eq} \sim {\cal
  O} (1)$ across stellar systems, if (for example) ${\cal
  R}_\text{planet} > 5 \, \text{AU}$ for a given Jupiter-like planet
compensates for PBHs with masses $m > 10^{14} \, \text{kg}$.

Given the form of the equilibrium capture number $N_\text{eq}$ in
Eq.~(\ref{Ncapture}), associated with PBH capture via three-body
interactions with a host star and a Jupiter-like planet with ${\cal
  R}_\text{planet}$, we may consider PBH orbits with longer semi-major
axes than the radius $R_\star$ of its host star, such as ${\cal
  R}_\text{pbh} \sim {\cal O} (1 \, \text{AU}) \simeq {\cal O}(10^8 \,
\text{km})$. The corresponding escape velocity at such distances is
$v_\text{esc} = \sqrt{ 2 G M_\star/ {\cal R}_\text{pbh} } \simeq 40
\, \text{km} \, \text{s}^{-1}$. For such PBH orbits, the GW waveforms are
strongly peaked at frequency $f_\text{peak} \simeq 1 / t_\text{orbit}
\simeq v_0 / {\cal R}_\text{pbh} \sim 10^{-7} \, \text{Hz}$. If we
approximate $A (f) \simeq A_\text{peak} \, \delta (f -
f_\text{peak})$, where $A_\text{peak} \equiv A (f_\text{peak})$, then
\begin{equation}
    \langle \vert \tilde{h}_+ (f) \vert^2 + \vert \tilde{h}_\times (f)
    \vert^2 \rangle \simeq A_\text{peak}^2 \, \delta (f -
    f_\text{peak}) ,
\label{hsquareAdelta}
\end{equation}
upon averaging over inclination angles $\eta$. At some cosmological
distance from a near-Earth detector, given by redshift $z$, the
measured frequency in the detector $f$ would be redshifted compared to
the frequency $f_\text{r}$ that an observer at rest near the source would
measure as $f = f_\text{r} / (1 + z)$. Given that current pulsar timing
arrays are sensitive to measured frequencies in the range $10^{-9} \,
\text{Hz} \leq f \leq 10^{-7} \, \text{Hz}$
\cite{NANOGrav:2023gor,EPTA:2023fyk,Reardon:2023gzh,Xu:2023wog}, we
therefore consider the contributions from a cosmic collection of such
PBH-star systems, with ${\cal R}_\text{pbh} \sim {\cal O} (1 \,
\text{AU})$.

For the peak amplitude $A_\text{peak}$, we again fix the star mass $M_\star =
M_\odot$ and $R = R_\odot$ and consider a typical comoving distance of
the PBH-stellar system to Earth to be $d = 3 \times 10^{13} \,
\text{km} = 1 \,\text{pc}$. Taking the dimensionless amplitude $\alpha_\text{max} \sim
     {\cal O} (10)$, as in Figs.~\ref{fig:Case1}--\ref{fig:Case3},
     then from Eq.~(\ref{hTTtypical}) we may estimate
\begin{equation}
    A_\text{peak} (z) \approx 10^{-30} \left( \frac{ m}{10^{20} \,
      \text{kg}} \right) \left( \frac{ 1 \, \text{AU}}{ {\cal
        R}_\text{pbh}} \right) \left( \frac{ 1 \,
      \text{pc}}{d_\text{com}} \right)\frac{1}{1 + z} ,
    \label{Apeakz}
\end{equation}
where the factor $1 / (1 + z)$ takes into account that the amplitude
near Earth from a source at comoving distance $d_\text{com}$ will be
redshifted by cosmic expansion.

A collection of independent PBH-star systems would contribute
incoherently to the SGWB intensity, with
\begin{equation}
    \vert A_\text{pbh}^\text{total} \vert^2 = \sum_i \vert A_i \vert^2
    \simeq \int N (z) \, \vert A_\text{peak} (z) \vert^2 \dd z,
\label{Atotal1}
\end{equation}
where $N(z)$ counts the number of such sources as a function of
redshift. To estimate $N(z)$, we write
\begin{equation}
N (z) \, \dd z \simeq N_\text{eq} \, f_\odot \, N_\odot \,
n_\text{gal} \, \dd V (z) ,
\label{Nz}
\end{equation}
where the capture rate $N_\text{eq}$ is given in Eq.~(\ref{Ncapture}),
$f_\odot \simeq 0.2$ is the fraction of Sun-like stars within the
Milky Way galaxy, $N_\odot \simeq 10^{11}$ is the total number of such
stars in the Milky Way galaxy, and $n_\text{gal} \simeq 2 \,
\text{Mpc}^{-3}$ is the number density of galaxies comparable to the
Milky Way within our local neighborhood (considering the Milky Way and
Andromeda to be the baseline). The volume factor out to redshift $z$
may be written in terms of the comoving volume element
\cite{Hogg:1999ad}
\begin{equation}
\dd V_\text{c} = R_\textsc{h} \frac{ d_M^2 (z)}{E(z)} \dd\Omega \, \dd z ,
\label{dVc}
\end{equation}
where $R_\textsc{h} \equiv c H_0^{-1} = 4350 \, \text{Mpc}$ is the present
value of the Hubble radius, $\dd\Omega$ is the solid angle element,
$d_\textsc{ct} (z)$ is the comoving transverse distance,
\begin{equation}
d_\textsc{ct} (z) = R_\textsc{h} \int_0^z \frac{ \dd z'}{E (z')} ,
\label{dM}
\end{equation}
and $E(z)$ is the dimensionless Hubble parameter, defined via $H (z) =
H_0 E(z)$, with
\begin{equation}
E^2 \equiv \Omega_\text{r} (1 + z )^4 + 
\Omega_\text{m} (1 + z )^3 +
\Omega_{\mathcal{K}} (1 + z )^2 +
\Omega_\Lambda,
\label{Ez}
\end{equation}
with $\Omega_\text{r}$, $\Omega_\text{m}$, $\Omega_{\mathcal{K}}$
and $\Omega_\Lambda$ the relative densities, w.r.t. $\rho_\text{c}$,
the critical density for a spatially flat universe, of radiation,
pressureless matter, spatial curvature and dark energy respectively.
The physical three-volume of a sphere centered on the Earth out to
redshift $z$ is related to $\dd V_\text{c}$ as $\text{Vol} (z) = (1 + z)^3
\int \dd V_\text{c}$. Combining these factors, using Eq.~(\ref{Ncapture}) for
$N_\text{eq}$ and the best-fit $\Lambda$CDM values for each $\Omega_i$
\cite{Planck:2018vyg} yields
\begin{equation}
\begin{split}
A_\text{pbh}^\text{total} &\approx 10^{-21} \left( \frac{ m}{10^{20}
  \, \text{kg}} \right)^{1/2} \left( \frac{{\cal R}_\text{planet}}{5
  \, \text{AU}} \right)^{3/4} \\ &\quad \quad \times \left( \frac{ 1
  \, \text{AU}}{ {\cal R}_\text{pbh}} \right) \left( \frac{ 1 \,
  \text{pc}}{d_\text{com}} \right) ,
\end{split}
\label{Atotal}
\end{equation}
upon integrating out to $z = 3$ to include the dominant epoch of
Sun-like star formation \cite{DunlopStarFormation}. For simplicity, we
have retained the typical dark-matter characteristics as in
Eq.~(\ref{Ncapture}), as well as keeping $M_\star = M_\odot$ and
$M_\text{planet} = M_\text{Jupiter}$.

This amplitude may be compared with the SGWB amplitude reported by the
NANOGrav collaboration, $A_\text{SGWB} \simeq 8 \pm 1 \times
10^{-15}$. (See Fig.~1 in Ref.~\cite{NANOGrav:2023hvm}.) The expected
contribution to the SGWB from a collection of PBH-star systems would
be comparable to the reported $A_\text{SGWB}$ if, for example, the
typical PBH mass were $m \sim 10^{26} \, \text{kg}$, the typical
planetary semi-major axis ${\cal R}_\text{planet} \sim 10^2 \,
\text{AU}$, and the typical comoving distance $d_\text{com} \sim
10^{-2} \, \text{pc}$. Although a typical PBH mass $m \simeq 10^{26}
\, \text{kg}$ is considerably larger than the asteroid-mass range
within which PBHs could constitute all of dark matter, present
observational bounds (such as microlensing) are consistent with
$f_\text{pbh} \simeq 0.1$ for $m \simeq 10^{26} \, \text{kg}$
\cite{Carr:2020xqk,Green:2020jor,Carr:2023tpt,Escriva:2022duf,Gorton:2024cdm}.
Moreover, $t_\text{dyn} \sim 10^{-1} (M_\odot / m )\, t_\text{orbit}
\sim 10^3 \, \text{yr}$, suggesting that the orbits of PBHs with these
masses would be stable against dynamical friction parametrically
longer than the time-scales $\Delta T \sim 1 / f_\text{min} \sim 10^2
\, \text{yr}$ to which present-day pulsar timing arrays are sensitive.


In addition to considering the amplitude of the contribution to the
SGWB, we may also consider the spectral index $\gamma$ associated with
such a collection of incoherent sources. The typical source that is
expected to contribute to the SGWB with frequencies to which pulsar
timing arrays are sensitive is binary inspirals of supermassive black
holes (SMBHs). These yield a spectral index $\gamma = 13/3$, which is
in tension with the value of $\gamma$ inferred by the NANOGrav
collaboration from their analysis of their 15-year dataset
\cite{NANOGrav:2023hvm,Sato-Polito:2025ivz}. To estimate the spectral
index $\gamma$ that would arise from an incoherent collection of
small-mass PBHs orbiting Sun-like stars, we follow
Ref.~\cite{Phinney:2001di} and parameterize the energy density in GWs
as
\begin{equation}
\Omega_\text{gw} (f) = \frac{ 1}{ \rho_\text{c}} \frac{ \dd \rho_\text{gw}
  (f)}{\dd \ln f} = \frac{ 8 \pi^4}{H_0^2} f^5 \frac{ \Phi (f)}{\Delta
  f} ,
\label{Omegagw1}
\end{equation}
with $\Delta f = 1 / T_\text{obs}$ is related to the observing
window. The function $\Phi (f)$ is typically parameterized as a power
law \cite{NANOGrav:2023hvm},
\begin{equation}
    \Phi (f) = \frac{ A^2}{12 \pi^2} \Delta f \left( \frac{f}{
      \text{yr}^{-1}} \right)^{- \gamma} \, \text{yr}^{3} ,
\label{Phidef}
\end{equation}
in terms of a GW waveform amplitude $A$. The energy density
$\Omega_\text{gw} (f)$ therefore scales as $\Omega_\text{gw} (f) \sim
f^{5 - \gamma}$ with some spectral index $\gamma$.

The energy density may be computed as \cite{Phinney:2001di}
\begin{equation}
\Omega_\text{gw} (f) = \frac{ 1}{ \rho_\text{c} c^2} \int \dd z \, 
\frac{N (z)}{ 1 + z} \left( f_\text{r} \frac{ \dd E_\text{gw}}{\dd f_\text{r}} \right)
\bigg\vert_{f_\text{r} = f (1 + z)} ,
\label{Omegagw2}
\end{equation}
with
\begin{equation}
    \frac{ \dd E_\text{gw}}{\dd f_\text{r}} = \frac{ 2 \pi^2 c^3}{G} \, d_\textsc{ct}^2
    (z) f^2 \langle \vert \tilde{h}_+ (f) \vert^2 + \vert
    \tilde{h}_\times (f) \vert^2 \rangle .
\label{dEdf}
\end{equation}
Making use of Eq.~(\ref{hsquareAdelta}) for a collection of
independent PBH-star systems then yields
\begin{equation}
\begin{split}
    \left( f_\text{r} \frac{ \dd E_\text{gw}}{\dd f_\text{r}} \right)
    \bigg\vert_{f_\text{r} = f (1 + z)} &\simeq \frac{ 2 \pi^2 c^3}{ G} d_\textsc{ct}^2
    (z) f^3 (1 + z)^3 \\ &\quad \times A_\text{peak}^2 (z) \,\delta
    \left[ f (1 + z) - f_\text{peak} \right] .
\end{split}
\label{dEdf2}
\end{equation}
Substituting into Eq.~\eqref{Omegagw1} suggests that 
\begin{equation}
\gamma\simeq 2
\label{gammapbh}
\end{equation}
for such a collection of sources. 

If the SGWB consisted of a combination of types of sources, such as
SMBH binary inspirals (with $\gamma = 13/3$) as well as bound PBH-star
orbits (with $\gamma \simeq 2$), then the weighted average
$\gamma_\text{avg}$ would fall closer to the central value $\gamma
\simeq 3.4 \pm 0.2$ inferred by the NANOGrav collaboration (see Fig.~1
of Ref.~\cite{NANOGrav:2023hvm}), if the corresponding amplitudes
$A_\text{SMBH}$ and $A_\text{pbh}^\text{total}$ for each type of
source were themselves comparable.


\section{Final Remarks}
\label{sec:Final}

In this work, we investigated the GW emission produced by a primordial
black hole orbiting a Sun-like star, whose mass distribution is
illustrated in Fig.~\ref{mass}. The equation governing the GW
production was written in such a way that all the physical parameters
characterizing the system appear as a coefficient of the dynamical
term, given by the second derivative of the quadrupole moment of the
system. Consequently, our findings can easily be applied to systems in
which the central star has a mass distribution with a similar
profile. The only need is to adjust the physical parameters to match
the values associated to the new system.

The shape of the orbits will naturally depend upon the initial
conditions, but the magnitude of the effect will mostly be given by
the coefficient appearing in the strain tensor in Eq.~\eqref{pbh2}.
For instance, an interesting system to investigate is one involving a
red dwarf (spectral type M), which is reported to be the most populous
type of star in the galaxy. Red dwarfs are generally less massive than
the Sun, with typical masses ranging from 0.08 to 0.6 $M_\odot$. Their
radii are smaller, too, typically spanning from 0.1 to 0.6
$R_\odot$. Regarding their mass-density distribution, they are claimed
to be denser near the center, as compared to the Sun. For a dwarf of
radius $0.1\, R_\odot$ and mass $0.1 \,M_\odot$, the central
mass-density is expected to be about $\rho_\textsc{c} \approx 500
\,\text{g}\,\text{cm}^{-3}$ \cite{Chabrier:1997vx}. Although a steeper
mass-density gradient is expected in these stars, its influence on the
magnitude of the GW amplitude does not significantly enhance the effect,
as the ratio $M_\star/ R_\star$ is approximately the same as that of the Sun,
$M_\odot/R_\odot$. However, given that their distance to the Earth is
much larger than that of our Sun, their GW strains will be several
orders of magnitude smaller. For instance, the closest red dwarf we
know is Proxima Centauri, approximately $4.02\times 10^{13} \,
\text{km} = 1.30 \, {\rm pc}$ away. The amplitude of a GW produced by a PBH of mass
$10^{20} \text{kg}$ orbiting such dwarf star, and measured near Earth,
would be $h^\textsc{tt} \approx 10^{-28}$.


In this work, we examined the particular case in which the observer is
placed orthogonally to the plane of the PBH's orbit. The
generalization to an arbitrary observer location is straightforward,
and it can be shown that it results in slightly different strains,
though they remain of the same order of magnitude. In particular, for
the case of a PBH that is orbiting the Sun in the same plane as the
Earth does, one of the strains could identically vanish, while the
other would remain unchanged. More general configurations, depending
on the observer's position relative to the orbital plane, would lead
to GW strains showing different patterns, when compared to the
orthogonal configuration.

Finally, we have computed expected GW strains arising from a
small-mass PBH orbiting a Sun-like star and considered whether such
systems could yield detectable GW signals. Whereas it is unlikely that
such an isolated system would yield a large enough amplitude to be
detected by \textsmaller{LISA}, we find regions of parameter space in which a large
collection of such systems, dispersed throughout the Universe over
much of cosmic history, could contribute in a measurable way to the
stochastic gravitational-wave background (SGWB). Moreover, the
spectral index expected for such a large collection of incoherent
sources would help alleviate the present tension with the recent
NANOGrav measurement, under the assumption that the signal arises
predominantly from the binary inspirals of supermassive black holes.

\begin{acknowledgments}
It is a pleasure to thank Bruce Allen, Josu Aurrekoetxea, Luc
Blanchet, Bryce Cyr, Valerio de Luca, Guillaume Faye, Peter Fisher, Evan Hall,
Benjamin Lehmann, Priya Natarajan, and Rainer Weiss for helpful
comments and discussion. V.~A.~D.~L. is supported in part by the
Brazilian research agency CNPq under Grant No. 302492/2022-4. L. R. S.
is supported in part by FAPEMIG under Grants No. RED-00133-21 and
APQ-02153-23. Portions of this work were conducted in MIT's Center for
Theoretical Physics and supported in part by the U.~S.~Department of
Energy under Contract No.~DE-SC0012567.
\end{acknowledgments}



%

\end{document}